\documentclass[12pt]{article}
\usepackage{graphicx}
\newcommand{\dnu}{\partial_{\nu}}
\newcommand{\Do}{\Omega}
\newcommand{\Ga}{\Gamma}
\newcommand{\Neu}{\mathcal{N}}
\newcommand{\sym}{\sigma}
\newcommand{\eps}{\varepsilon}
\newcommand{\lmax}{l_{\rm max}}
\newcommand{\nmax}{n_{\rm max}}
\newcommand{\mmax}{m_{\rm max}}
\newcommand{\ha}{H_0^{(1)}}
\newcommand{\dst}{\displaystyle}
\newcommand{\bR}{\mathbf{R}}

\newcommand{\syml}{\sym_{\mathrm{lim}}}
\def\rem#1{} 
\begin{document}
\begin{center}
{\Large 
Numerical study of high frequency asymptotics \\
of the symbol of the Dirichlet-to-Neumann operator \\[0.7ex]
in 2D diffraction problems
}
\\[0.5cm]
{\bf Margo~Kondratieva and Sergey~Sadov}
\\[0.5cm]
{
Department of Mathematics and Statistics \\ 
Memorial University of~Newfoundland \\
St.\ John's NL, A1C 5S7, Canada. \\[2ex]
mkondra@math.mun.ca, $\;$ sergey@math.mun.ca
}
\end{center}

\vskip 0.3cm
\hangindent=0.1in
\noindent \hspace*{0.1in}
{\small
{\bf Abstract.}
A high-frequency asymptotics of the symbol of the
Dirichlet-to-Neumann map, treated as a periodic 
pseudodifferential operator, in 2D diffraction problems 
is discussed. Numerical results support
a conjecture on a universal limit shape of the symbol.
}

\vskip 0.2cm
\hangindent=0.1in
\noindent \hspace*{0.1in}

\section{Introduction}

The classical Kirchhoff Approximation (KA) in diffraction theory is an asymptotic relation
between the Dirichlet and Neumann data of a solution of an exterior boundary value problem for
the Helmholtz equation with frequency parameter $k\gg 1$.
The KA is sensitive to the assumption of positive curvature of the boundary
and its accuracy deteriorates in the presence of flattening
regions \cite{Taylor}. Numerical methods for high-frequency
problems have attracted much attention lately -- see e.g.~\cite{Bruno}.
In an attempt to include small and vanishing
curvatures uniformly in an asymptotic theory, we propose to study
high-frequency asymptotic properties of the Dirichlet-to-Neumann
(DtN) operator rather than those of an individual solution.
In \cite{dd03}, we conjectured that the pseudodifferential symbol of the DtN operator,
appropriately scaled, tends to a simple universal function as
$k\to \infty$. Here we report results of a more detailed numerical study.
The results support the said conjecture in the case of a convex scatterer.
In a non-convex case, we observe a deviation from the universal limit function
in a narrow range of it's argument values.

\section{The DtN operator and the Limit Shape Hypothesis}

Consider the Helmholtz equation
$\Delta u+k^2u=0$ in the exterior of a simply connected
bounded domain $\Do \subset \bR^2$
with smooth boundary $\Ga$.
Given a function $f$ on $\Ga$ of a certain regularity \cite{Col1,Ramm},
the Dirichlet problem
$\,u|_{\Ga}= f\,$ has a unique solution $u$
satisfying the Sommerfeld radiation condition
$\,\partial_r u-iku=o(r^{-1/2})$ as $\,r\to \infty$.
The normal derivative $\,g=\dnu u|_{\Ga}\,$  
is a function of known regularity on $\Gamma$.
The map $\Neu:\,f\to g$ is called the Dirichlet-to-Neumann (DtN) operator.

Let $s$ be the arclength parameter on
$\Ga$, and $L$ the length of $\Ga$.
Set $\phi=2\pi s/L$.
The Dirichlet and Neumann data $f$ and $g$
in the above diffraction problem
are $2\pi$-periodic functions of variable $\phi$.
Let $\,f(\phi)=\sum \hat{f}(n) e^{in\phi}$ be the Fourier series of $f$.
Write $\Neu$ as a periodic pseudodifferential operator (PPDO)
\cite{Agr,Saranen}
\begin{equation}
\label{ppdo}
 g(\phi)=\Neu f(\phi)\,=
 \,\sum_{n=-\infty}^{\infty} \sym(\phi,n)\,\hat{f}(n) e^{in\phi}.
\end{equation}
The function $\;\sym(\phi,n)=e^{-in\phi}\,\Neu e^{in\phi}$
is called the {\it symbol} of $\Neu$.

The operator $\Neu$ depends on the boundary $\Ga$ as well as on
the frequency $k$. We reflect this in notation of the symbol
by writing $\,\sym(\phi,n)=\sym_{\Ga}(\phi,n;\,k)$.
In \cite{dd03} we noted an universal ($\Ga$-independent)
high-frequency asymptotic behaviour of the symbol as a function of variable
$\,\dst \xi=\xi(n,k)=\frac {2\pi n}{L k}\,$.
Define
$$
\syml(\xi)\;=\;\left\{\;\begin{array}{lr}
i\sqrt{1-\xi^2}, & \quad |\xi|<1, \\
-\sqrt{\xi^2-1}, & \quad |\xi|\ge 1.
\end{array}\right.
$$

\noindent
{\bf Hypothesis 1.}
{\it For any $\eps>0$ and any $\xi_*>1$, there exists
 $k_*>0$ such that
\begin{equation}
\label{eq6}
\left|k^{-1}\;\sym_{\Ga}(\phi,n;\,k)
-\syml\left(\xi(n,k)\right)\right|\le \eps
\end{equation}
whenever $\,k\ge k_* \,$ and $\,|\xi(n,k)|\le\xi_*$.
}

\medskip
Here are some theoretical arguments in favour of Hypothesis~1.

\begin{enumerate}
\item 
The statement holds if $\Ga $ is a circle of any radius
\cite{dd03}.

\item 
If $\,|\xi(n,k)|>1$, then the inequality
(\ref{eq6})
can be established by constructing an asymptotic WKB solution, as
pointed out by L.~Friedlander (Univ.~of Arizona), personal communication,
February 2004.

\item 
A simple if not completely rigorous argument shows that
the hypothesis is consistent with KA for a convex domain
\cite{dd03}.
\end{enumerate}

Yet we admit that Hypothesis~1 may be true for some classes of boundary
curves and false for others. To restore {\em status quo}\
with numerical experiment, we formulate a somewhat weaker Hypothesis 2 below.

Note that the symbol $\sigma_{\Ga}(\phi,n;\,k)$
generally depends on $\phi$ (except when $\Ga$ is a circle),
while the limit function is $\phi$-independent.
So we are trying to approximate the DtN operator by a shift-invariant
PPDO. It can only be possible if the Fourier series of the symbol
in $\phi$ asymptotically reduces to a single constant term. Put
(omitting the subscript $\Ga$ in the right-hand side)
\begin{equation}
\label{eq8}
\sym_{\Ga}(\phi,n;\,k)=
\hat\sym_0(n;\,k)+\hat\sym_{\pm 1}(n;\,k)e^{\pm i\phi}
+\hat\sym_{\pm 2}(n;\,k)
e^{\pm  2 i\phi}+\cdots  .
\end{equation}
We shall compare the {\it mean symbol}\
$\,\hat\sym_0(n;\,k)
=(2\pi)^{-1}\int\nolimits_0^{2\pi}\sym_{\Ga}(\phi,n;\,k)\,d\phi\,$
to the limit function and watch whether the
$l_2$-norm $||\hat \sym'(n;\,k)||$ of a bi-infinite vector
formed by the rest of Fourier coefficients (\ref{eq8})
is relatively small. Recall:
\begin{equation}
\label{eq9}
||\hat \sym'(n;\,k)||^2={\sum_{m\ne 0} \left|\hat\sym_{m}(n;\,k)
\right|^2}.
\end{equation}

\noindent
{\bf Hypothesis 2.}
{\it For any boundary curve $\Ga$
and any given $\xi_*>1$, $\eps>0$, and $\delta>0$, there exists $k_*>0$
such that if $\,k\ge k_* \,$ and
$\,|\xi(n,k)|\le \xi_*$, then

\medskip
\begin{enumerate}
\item 
the shape of the mean symbol $\hat\sym_0$ of the DtN operator follows
that of $\syml$:
\begin{equation}
\label{eq6b}
|k^{-1}\;\hat\sym_0(\xi(n,k);\,k)-\syml\left(\xi\right)|\le \eps;
\end{equation}

\item 
the remaining Fourier coefficients of the symbol are collectively small:
\begin{equation}
\label{eq6c}
k^{-1}||\hat \sym'(n,k)||\le\eps,
\end{equation}
%
if $\,{\rm distance}\,(\xi(n;k),\, I)>\delta$. Here $I$ 
is either the empty set or a certain ``exceptional'' set
determined by the curve $\Ga$.

\end{enumerate}
}

Note that Hypothesis 1 implies Hypothesis~2 with
$I
=\emptyset$. The parameter $\delta$ in Hypothesis 2 is introduced 
to account for a non-uniform convergence near 
$I$ when
$I$ is nonempty.
Note also that in this paper we require $\Ga$ to be a smooth curve,
but there exist numerical results supporting validity of the statement
for domains with corners.

\section{Methodology of numerical verification}
To test the hypothesis numerically, we use known sample solutions
satisfying the Helmholtz equation (HE) in the exterior domain
$\bR^2\setminus \overline{\Do}$
and the radiation condition (RC), and compute Fourier coefficients
of the Dirichlet and Neumann data. Solutions of HE
in $\bR^2\setminus(0,0)$ with wavenumber $k$ and satisfying
RC are spanned by the Hankel functions
$H_m^{(1)}(kr)$, $m=0,1,\cdots$, $\,r=|\vec r|=\sqrt{x^2+y^2}$.
The origin can be viewed as an emitter, or source.
Now, by taking fictitious sources at arbitrary locations
$\vec S\in \Do$, the family $\,H_m^{(1)}(k|\vec r-\vec S|)\,$
of sample solutions in $\bR^2\setminus \overline{\Do}\,$ is constructed.
For the verification procedure one can use a countable
sub-family with linear combinations dense in the
space of solutions. In this work, we use $\ha$-solutions
with sources near the boundary $\Ga$ and approximately equidistributed
along $\Ga$. A possibility to represent an arbitrary solution
of HE$+$RC in the form of a single layer potential
(provided $k^2$ is not an interior eigenvalue \cite[\S{}3.2.1]{Ned3}),
justifies this choice. An extreme opposite
possibility is to choose a family of $H_m^{(1)}$-solutions, $m=0,1,2,\cdots$,
with fixed source. It needs the Rayleigh hypothesis for domain $\Do$ to
hold, which is true, for example, if $\Ga$ is an ellipse with eccentricity
$e<1/\sqrt{2}\,$ \cite{Ray}.

Let us first describe a procedure used in \cite{dd03}.
Take a uniform partition $\{\vec P_l\}$, $l=1,2,\ldots, \lmax$
of the curve $\Ga$. Evaluate a sample solution
$\ha(k|\vec r-\vec S|)$ and its normal derivative
at the points $\vec r=\vec P_l$ to obtain the vectors
$f_l$ and $g_l$ of size $\lmax$.
Then compute the discrete Fourier transforms
and consider their truncations $\hat f(n)$, $\hat g(n)$,
$|n|\le\nmax$.
Find the ratio $\,\tilde \sym(n)=\hat f(n)/\hat g(n)\,$
and compare $\,k^{-1}\tilde \sym(n)\,$ to
$\,\syml\left(\frac{2\pi n}{Lk}\right)\,$
to verify Hypothesis 1.
Typically in our examples $Lk\sim 10^2\div 10^3$;
we chose $\lmax=2^{12}\div 2^{24}$, and $\nmax\approx 3 Lk$.
Higher Fourier coefficients are vanishingly small,
that is why we cut them off.

In more detail, let $\vec r(\phi)$,
$\phi\in [0,2\pi]$ be the parametrization of $\Ga$ by the normalized
arclength $\phi=2\pi s/L$. Put $\,\phi_l=2\pi l/\lmax\,$ and
$\,P_l=\vec r(\phi_l)$. Then
$$
f_l=\ha(k|\vec r(\phi_l)- \vec S|),
\qquad
g_l=-\eps_l\, k\left[1-\left(r'(\phi_l)\right)^2
\right]^{1/2}
\,H_1^{(1)}(k|\vec r(\phi_l)- \vec S|).
$$
Here $r(\phi)=|\vec r(\phi)|$ and
$\eps_l=(-1)^{p_l}$, where $p_l$ is the number of intersections
of $\Ga$ with the open interval $(S P_l)$.
Note  that $p_l\equiv 0$ and $\eps_l\equiv 1$ if $\Do$ is convex.

Our judgement about validity of Hypothesis 1 in \cite{dd03}
was based on the outlined procedure, where we effectively
kept over the mean symbol $\hat\sym_0(n;\,k)$ only. But this is not enough.
Let us engage in the study of components of the vector $\hat \sigma'$,
see (\ref{eq9}), -- apart from $\hat\sym_0(n;\,k)$.
Now we take several sources, $S_1,\ldots,S_J$ at once.
Assume, for the sake of symmetry, that $J$ is odd, $J=2\mmax+1$.
Denote by $f_l^j$ and $g_l^j$ the data of the solution with
source at $S_j$, and by $\hat f^j(n)$, $\hat g^j(n)$
the corresponding components of the (truncated) discrete Fourier transforms.
The following relations follow from
(\ref{ppdo}),(\ref{eq8}): for every $j=1,\ldots,J$
\begin{equation}
\label{symsys}
\hat g^j(n)\,=\,\sum_m\hat f^j(n-m)\,\hat\sym_{m}(n-m;\,k).
\end{equation}
Reduce the infinite summation
to a finite number of terms keeping only the components
$\hat\sym_m(\,\cdot\,;k)$ with $|m|\le \mmax$.
For example,  if $J=3$, then for each $n=-\nmax,\ldots,\nmax\,$
after cut-off we get a linear system 
of three equations with three
unknowns $\hat \sym_0(n;k)$, $\hat \sym_{\pm 1}(n;k)\,$:
$$
\hat g^j(n)= \hat f^j(n+1)\hat\sym_{-1}(n+1)
+\hat f^j(n)\hat\sym_{0}(n)
+\hat f^j(n-1)\hat\sym_{1}(n-1),
\quad j=1,2,3.
$$
Solving all obtained
systems, we approximately find $\hat\sigma_m(n)$ for (at least)
$|n|\le\nmax-\mmax$ and $\,|m|\le \mmax$.
Now the left-hand sides of the inequalities (\ref{eq6b}), (\ref{eq6c})
can be evaluated; of course, summation in (\ref{eq9}) is restricted
to $|m|\le \mmax$.

\noindent
\begin{figure}[h]
\centerline{
\begin{picture}(162,139)
\put(19,8){\includegraphics{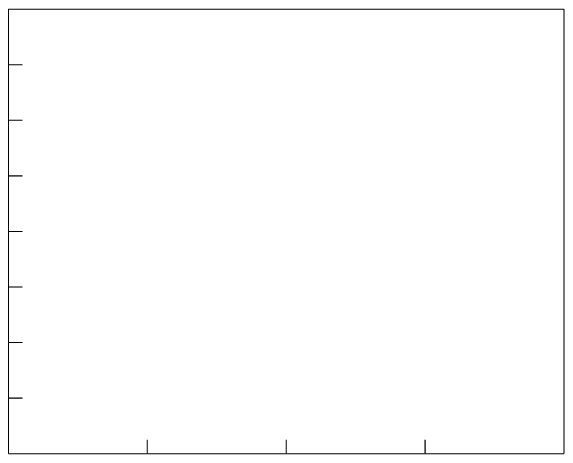}}
\put(19,24){\includegraphics{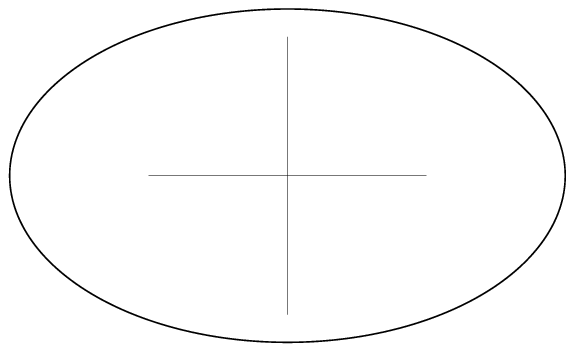}}
\put(9,71){\small 0}
\put(3,119){\small 0.6}
\put(0,23){\small -0.6}
\put(97,0){\small 0}
\put(135,0){\small 0.5}
\put(52,0){\small -0.5}
\put(0,0){(a)}
\end{picture}
\hspace{60pt}
\begin{picture}(162,139)
\put(0,8){\includegraphics{frame124}}
\put(20,13){\includegraphics{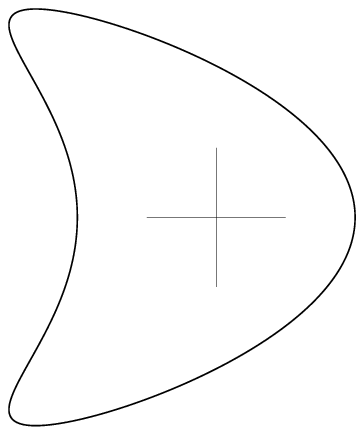}}
\put(-10,71){0}
\put(-16,119){\small 1.2}
\put(-19,23){\small -1.2}
\put(78,0){\small 0}
\put(120,0){\small 1}
\put(35,0){\small -1}
\put(-17,0){(b)}
\end{picture}
}
\caption{Test domains: (a) Convex (ellipse), (b) Non-convex (kite
\cite{Col2})}
\end{figure}

\section{Results and discussion}

We present results for two symmetric domains shown on Fig.~1:
the ellipse $x(t)=\cos t$, $y(t)=0.6\sin t$,
and a non-convex ``kite'' 
$\; x(t)=\cos t+0.65 \cos 2t-0.65$,
$y(t)=1.5\sin t$.  
If $\phi=0$ at the right $x$-intercept of $\Ga$, then due to symmetry
$\,\hat\sym_m(-n;k)=\hat\sym_{-m}(n,k)$ and it suffices
to study the symbols for $n\ge0$. 

The real and imaginary parts
of the rescaled mean symbol $k^{-1}\hat\sigma_0(n,k)$ are compared to
the limit curves on Fig.~2,~3.
Here $\xi=\xi(n,k)$ as defined in Sect.~2.
The parameters are: frequency $k=200$; number of sources $J=201$.
The kite's curves exhibit some roughness when $\xi\in (0.8,1)$.

\noindent
\begin{figure}[!hbt]
\centerline{
\begin{picture}(162,139)
\put(17,8){\includegraphics{frame124}}
\put(17,24){\includegraphics{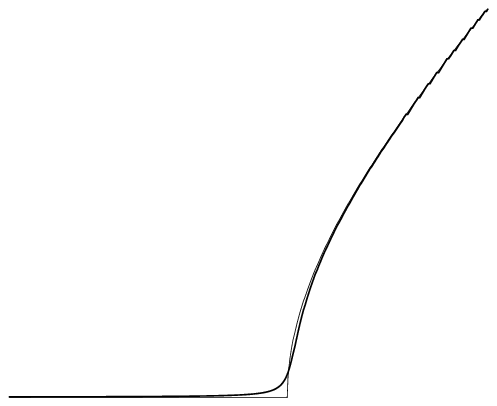}}
\put(7,102){\small 1}
\put(7,23){\small 0}
\put(95,0){\small 1}
\put(133,0){\small 1.5}
\put(50,0){\small 0.5}
\put(0,0){(a)}
\put(58,125){\small $k=200$}
\put(58,111){\small ellipse}
\end{picture}
\hspace{60pt}
\begin{picture}(162,139)
\put(0,8){\includegraphics{frame124}}
\put(0,21){\includegraphics{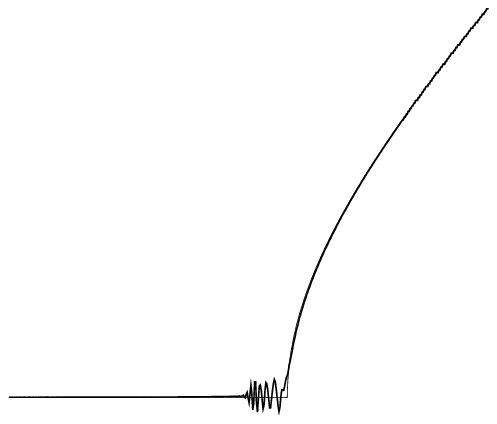}}
\put(-10,102){\small 1}
\put(-10,23){\small 0}
\put(78,0){\small 1}
\put(116,0){\small 1.5}
\put(33,0){\small 0.5}
\put(-17,0){(b)}
\put(42,125){\small $k=200$}
\put(42,111){\small kite}
\end{picture}
}
\caption{ $\;-\mathrm{Re}\,\hat\sym_0(n;k)/k\;$ and
$\;-\mathrm{Re}\,\syml(\xi)\;$ vs $\;\xi =\xi(n,k)$
}
\end{figure}

\noindent
\begin{figure}[!hbt]
\centerline{
\begin{picture}(162,123)
\put(17,8){\includegraphics{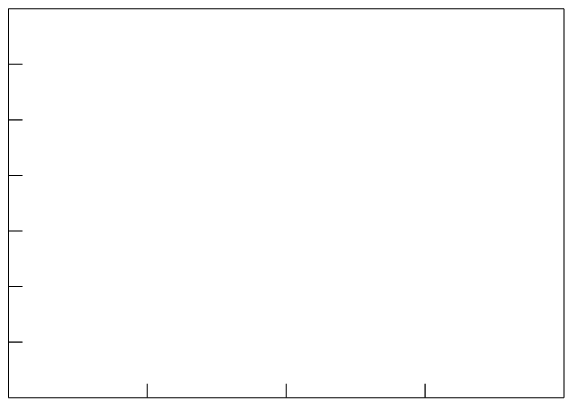}}
\put(17,24){\includegraphics{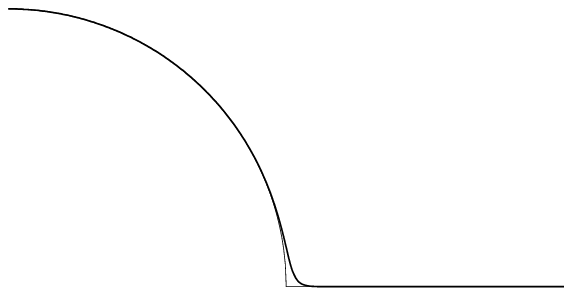} }
\put(7,102){\small 1}
\put(7,23){\small 0}
\put(95,0){\small 1}
\put(133,0){\small 1.5}
\put(50,0){\small 0.5}
\put(0,0){(a)}
\put(98,104){\small $k=200$}
\put(98,90){\small ellipse}
\end{picture}
\hspace{60pt}
\begin{picture}(162,123)
\put(0,8){\includegraphics{frame3}}
\put(0,24){\includegraphics{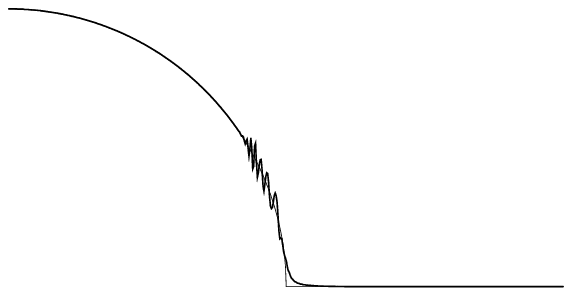}}
\put(-10,102){\small 1}
\put(-10,23){\small 0}
\put(78,0){\small 1}
\put(116,0){\small 1.5}
\put(33,0){\small 0.5}
\put(-17,0){(b)}
\put(82,104){\small $k=200$}
\put(82,90){\small kite}
\end{picture}
}
\caption{$\;\mathrm{Im}\,\hat\sym_0(n;k)/k\;$
and $\;\mathrm{Im}\,\syml(\xi)\;$ vs $\;\xi=\xi(n,k)$
}
\end{figure}

\vspace{-1.2\medskipamount}
\noindent
\begin{figure}[!hbt]
\centerline{
\begin{picture}(162,139)
\put(19,8){\includegraphics{frame124}}
\put(19,8){\includegraphics{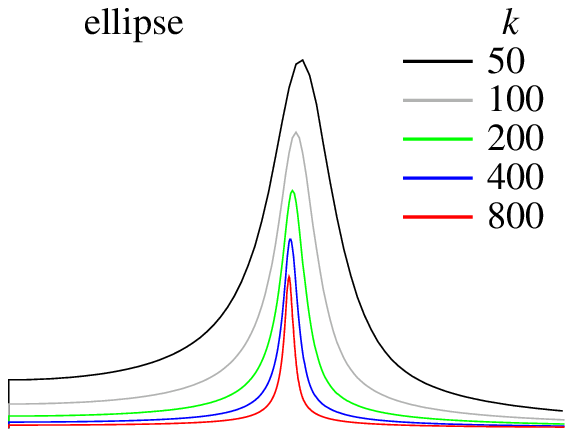}} 
\put(4,87){\small 0.2}
\put(0,119){\small 0.28}
\put(97,0){\small 1}
\put(135,0){\small 1.5}
\put(52,0){\small 0.5}
\put(0,0){(a)}
\end{picture}
\hspace{60pt}
\begin{picture}(162,139)
\put(0,8){\includegraphics{frame124}}
\put(0,9){\includegraphics{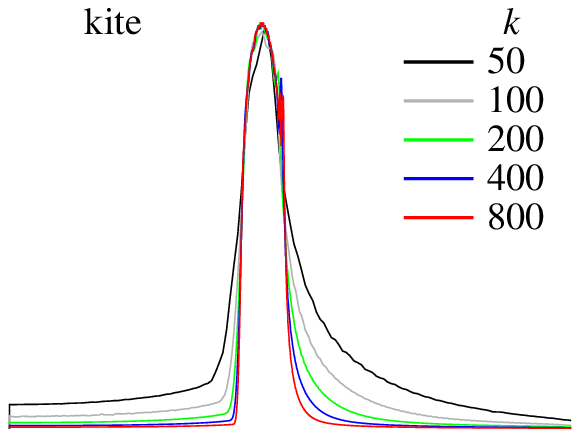}} 
\put(-16,87){\small 1.0}
\put(-16,119){\small 1.4}
\put(78,0){\small 1}
\put(116,0){\small 1.5}
\put(33,0){\small 0.5}
\put(-17,0){(b)}
\end{picture}
}
\caption{Violation of symbol shift-invariance:
$\;k^{-1}||\hat\sym'(n;k)||\,$ vs $\xi(n,k)$
}
\end{figure}

Fig.~4 shows the left-hand side of the inequality
(\ref{eq6c}) vs $\xi(n,k)\,$ for
$\,k=50$, 100, 200, 400, 800.
The value of $J$ was always set equal to $k+1$.
In the case of ellipse, the norm shrinks to naught as $k$ grows.
It isn't quite so for the kite. The peak over the interval
$(0.8,1)$ stays steady. 
In the frameworks of Hypothesis 2, we say that  
the exceptional set $I$ is empty for the ellipse, 
though the convergence near $|\xi|=1$ is much slower
than away from $|\xi|=1$. 
The set $I$ for the kite is apparently contained in the union 
$(-1,-0.8)\cup(0.8,1)$. 

\bigskip
{\bf Computational note.}\@
Computation of the Fourier coefficients 
$\hat\sym_m(n;k)$  of the symbol requires solution of truncated systems 
(\ref{symsys}).
If the cutoff subscript is rather large, one has to take trouble
to ensure that Fourier coefficients $\hat f^j (n-m)$ are not
vanishingly small.
To this end, the sources should be placed close to the boundary,
preventing the Dirichlet data of sample solutions from being ``too
smooth''.
The reported results are obtained with sources located at the
distance from about $10^{-2}$ to $10^{-3}$
(for larger values of $k$) from $\Gamma$.



\end{document}